\documentclass[prl,11pt, superscriptaddress, notitlepage]{revtex4-1}
\usepackage{graphicx}%
\usepackage{wrapfig}
\usepackage{float}
\usepackage[small, justification=justified]{caption}
\usepackage{subeqnarray,epsfig,amsmath}

\providecommand{\e}[1]{\ensuremath{\times 10^{#1}}}

\begin{document}

\title{Effects of Substrate Defects on Lipid Bilayer Compression Dynamics}

\author{Austin Fergusson}
\affiliation{Department of Engineering Science \& Mechanics, Virginia Tech, Blacksburg, VA 24061}
\author{Ravi Kappiyoor}
\affiliation{Department of Engineering Science \& Mechanics, Virginia Tech, Blacksburg, VA 24061}
\author{Ganesh Balasubramanian}
\affiliation{Department of Mechanical Engineering, Iowa State University, Ames, IA 50011}
\author{Ishwar K. Puri}
\affiliation{Department of Mechanical Engineering, McMaster University, Hamilton, Ontario L8S4L7, Canada}
\author{Douglas P. Holmes}
\email{dpholmes@vt.edu}
\affiliation{Department of Engineering Science \& Mechanics, Virginia Tech, Blacksburg, VA 24061}

\date{\today}%
\begin{abstract}
{\em In vivo} and {\em in vitro} lipid bilayers are commonly supported by subcellular structures, particles, and artificial substrates. Deformation of the underlying structure can lead to large, localized deformations as the bilayer deforms to avoid stretching. In this work, we consider the effect of defects within the underlying substrate by simulating different bilayers supported by continuous and nanoporous substrates. We show that the bilayer behavior greatly depends on strain rate, and that substrate defects may contribute to the formation of nanotubes for compressed substrates.
\end{abstract}
\maketitle

\section{Introduction}

\begin{wrapfigure}{r}{82.55mm}
\begin{center}
\vspace{-8mm}
\resizebox{0.44\columnwidth}{!} {\includegraphics{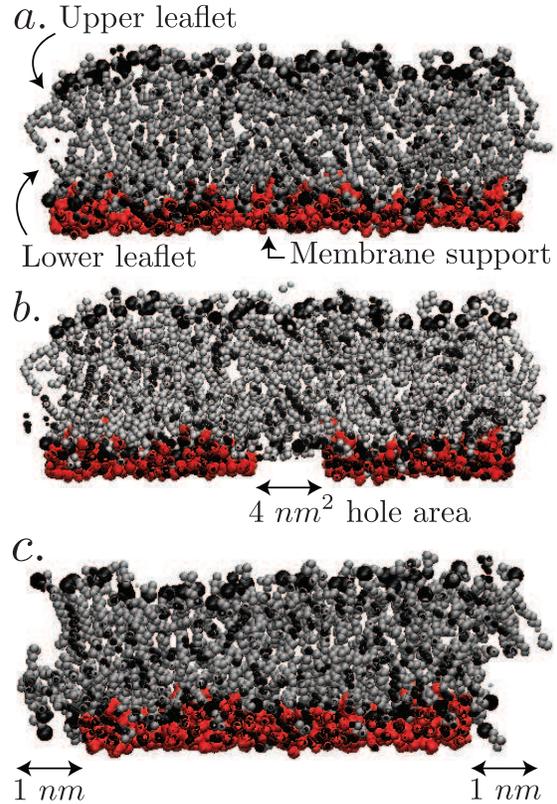}}
\end{center}\vspace{-6mm}
\captionsetup{font=footnotesize}
\caption{The simulated system contains a DPPC bilayer consisting of polar head groups (black) and hydrophobic tails (grey), a supporting substrate (red), and water layers above the membrane and below the substrate.  Cross sections  of the supported lipid bilayer and substrate ($a.$) with a center defect ($b.$) and edge defect ($c.$).}
\label{fig-1}\vspace{-5mm}
\end{wrapfigure} 


Due to their ability to provide electrical insulation, incorporate receptor proteins and suppress nonspecific ligand binding, supported lipid bilayers (SLBs) are used in biosensors~\cite{Wu2001}. Their mechanical properties have also been extensively investigated~\cite{Sackmann1996, Staykova2011, Staykova2013} Transmembrane phenomena, such as the ion channel function~\cite{Reimhult2008}, have been explored by suspending SLBs across nanopores~\cite{Kresak2009, Nellis2011, Simon2007}. The suspended SLB regions minimize substrate effects on system behavior~\cite{Reimhult2008}, and provide an ideal environment for these investigations. Applications of SLBs commonly assume that the supporting substrate is continuous and defect-free~\cite{Varma2012}. However, this assumption is unrealistic {\em in vivo}. The lipid bilayer that forms the cell membrane is supported by an F--actin mesh that contains large gaps between its fibers, suggesting that a supporting scaffold model would be more appropriate~\cite{Richard2008}. While experiments characterized the influence of the cytoskeleton on static membrane properties~\cite{Feng2009} and prior computational works by Xing {\em et al.}~\cite{Xing2008, Xing2009} and Lin {\em et al.}~\cite{Lin2012} investigated the changes in static membrane properties related to confinement by a supporting substrate, the effects of nanoscale defects in the substrate on membrane dynamics are still poorly understood. We investigate these dynamics through molecular dynamics (MD) simulations of the phospholipid dipalmitoylphosphatidylcholine (DPPC) (Fig.~\ref{fig-1}a). Two types of defects are simulated: (1.) a system with a hole in the center of the substrate (hereafter referred to as the center-defect substrate) that is shown in Fig.~\ref{fig-1}b, and (2.) a system with the substrate edges removed (referred to as the edge-defect substrate), shown in Fig.~\ref{fig-1}c. We also examine the effect of a third defect, namely, substrate protrusions, on SLB dynamics. These three scenarios are investigated by compressing the systems at a constant strain rate, which allows us to observe how the SLB accommodates the induced stress, both in the presence and absence of substrate defects. 

\section{Computational Methods}

\begin{wrapfigure}{r}{82.55mm}
\begin{center}
\vspace{-8mm}
\resizebox{0.44\columnwidth}{!} {\includegraphics{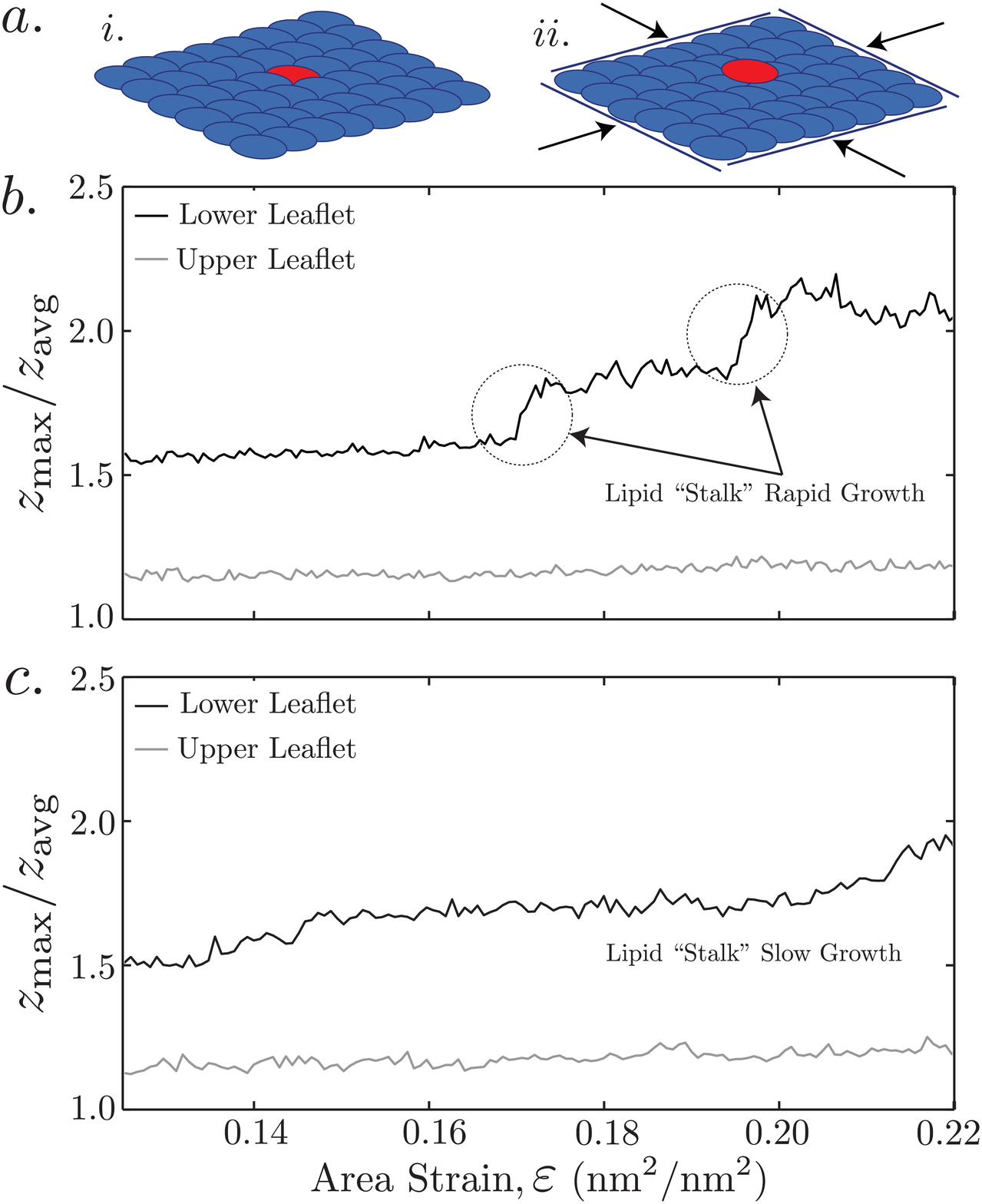}}
\end{center}\vspace{-6mm}
\captionsetup{font=footnotesize}
\caption{($a.$) A schematic of lipid head groups of the lower leaflet $i.$ which is pushed out of plane of the remaining head groups, $ii.$ which we describe as ÒbucklingÓ, upon reading a critical biaxial compression. The ratio of the maximum and average lipid head group heights versus compressive area strain plots for the defect free case. ($b.$) The 8\e{-5} nm/ps compression rate plot. The graph displays distinct jumps representing the elongation of the lipid stalk in the lower leaflet. ($c.$) The 1\e{-4} nm/ps compression rate plot, which shows a smoother, continuous growth process of the lipid stalk in the lower leaflet. The initial elongation of the ``stalk'' begins with a lower compressive area strain at the faster compression rate.} \vspace{-5mm}
\label{fig-2}
\end{wrapfigure}

The simulations are conducted with GROMACS~\cite{Hess2008} to perform: ($i.$) energy minimization, ($ii.$) a canonical $NVT$ (constant volume and constant temperature) ensemble simulation over 5 ns with the temperature controlled at 323 K by rescaling the molecular velocities every 1 picosecond (ps), and ($iii.$) an $NPT$ (constant pressure and temperature) ensemble based simulation over 13 ns using a Nose--Hoover thermostat~\cite{Nose1983, Hoover1985} with a coupling time of 1 ps, and a Parrinello-Rahman barostat~\cite{Parrinello1981, Klein1983} with a coupling time of 5 ps~\cite{Nose1983, Hoover1985, Parrinello1981, Klein1983, Bussi2007}. The dimensions of the system are: 12.584 nm x 12.633 nm x 10.474 nm. All simulations employed a time step of 0.002 ps. We used a simulation temperature and pressure of 323 K and 1 bar, respectively. The temperature was above the gel transition temperature of DPPC, and is commonly used in the literature~\cite{Nagle1993, Heller1993, Leekumjorn2007, Nagle1988, Nagle2000, Kodama1982}. Visualization of the structures is performed using VMD~\cite{Humphrey1996}.

The initial structure of DPPC is acquired from a previous study~\cite{Tieleman1996}: 128 lipid molecules arranged into two monolayers -- referred to as leaflets -- with 64 lipids each, as shown in Fig.~\ref{fig-1}. We use a modified GROMACS 53a6 force field including Berger lipid parameters for our simulations~\cite{Tieleman1996, Tieleman1997, Marrink1998}. The systems are simulated in bulk water (modeled using the SPC potential) with periodic boundary conditions~\cite{Berendsen1981}. The SETTLE algorithm is used to hold the geometry of each solvent water molecule fixed~\cite{Miyamoto1992}. The geometry of each water molecule in the substrate is held fixed using the SHAKE algorithm~\cite{Ryckaert1977}. Coulombic and van der Waals interactions have a cut-off distance of 1.2 nm~\cite{Berendsen1981, Darden1993}. Long-range charge interactions are computed using a Particle Mesh Ewald (PME) summation scheme~\cite{Darden1993}. After equilibration, the area per lipid head group is 0.62 nm$^2$, which is in good agreement with previous work that reports this value as $0.629 \pm 0.013$ nm$^2$~\cite{Nagle1996}. We construct a larger system by replicating the equilibrated structure along the $x$ and $y$--directions. The larger structure is desirable for compression simulations to ensure that the $x$ and $y$--dimensions of the lipid bilayer are several times larger than the substrate defects. This structure is subjected to another NPT equilibration that is simulated for 2 nanoseconds using the same thermostat and barostat as before. Once the density and temperature data converge, the system is ready for data production runs.

The supporting substrate consists of water molecules within $\sim$2 nm from the bottom of the lipid bilayer. They are restrained from translational motion by applying an energy penalty to each atom. This penalty must be overcome to produce movement. We refer to the supporting substrate as the ``water slab''. The water slab restricts the ability of the bilayer to adjust during compression. Although the choice of support differs from typical experiments with SLBs, the simulations are still qualitatively comparable due to the polar nature of the support. To create the desired nanoscale defects in the substrate, the translational constraints on specific water molecules are removed, thereby allowing the bilayer to deform through these regions. The center-defect substrate has a 4 nm$^2$ hole in the center while the edge-defect substrate has 1 nm of the substrate removed along the outer edges.

Once this structure equilibrates, we conduct compression simulations. The pressure control during all compression simulations is coupled along the $x$ and $y$--directions. The corresponding control in the $z$--direction is decoupled to allow system deformations that are normal to the plane of the bilayer to differ from those in the tangential directions. The system is compressed along the plane of the membrane (i.e. in the $x$ and $y$--directions) at four deformation rates: 8\e{-5}, 1\e{-4}, 2\e{-4}, and 3\e{-4} nm/ps. As a control, we also simulate a defect-free substrate at compression rates of 8\e{-5} and 1\e{-4} nm/ps. Simulations are run until the bilayer reaches an area strain of $\sim30$\%. The simulations times for the different compression rates are (fastest -- slowest): 10 ns, 15 ns, 30 ns, and 35 ns. Analysis becomes difficult thereafter due to the amount of bilayer deformation. We make the assumption that no lipid molecules switch leaflets because the flip-flop rate for DPPC is on the timescale of hours~\cite{Sapay2009} which is orders of magnitude longer than the simulations performed in this work.

The mean curvature of both leaflets is calculated by applying a finite difference approximation of the Laplacian to the surface that is interpolated from the positions of the lipid head groups. We consider the compressive area strain given by
\begin{equation}
U_A=\frac{A_0-A_{new}}{A_0}
\end{equation}
where $A_0$ represents the lateral area of the membrane in its uncompressed state and $A_{new}$ is the same at a specific time step. Since this work only examines membrane compression, $U_A$ is always positive.

\section{Results and Discussion}

In each of the defect-free simulations, the SLB remains planar until it experiences a critical area strain when the lower leaflet buckles. Buckling, which occurs at the corners of the simulation box, is considered to arise when a lipid is forced out-of-plane (Fig.~\ref{fig-2}a). Compression causes the substrate molecules to pile up at the corners of the box. This protrusion promotes deformation. The compression dynamics of the SLB are biased toward delamination at the promoter site. We postulate that membrane delamination at these sites represents the initial stages of lipid tether formation.

Further compression drives additional lipid head groups in the lower leaflet upwards into the interior of the bilayer, leading to the growth of a lipid ``stalk'' in the buckling region. We monitor the ratio of the $z$--coordinate of the highest lipid head group to the mean $z$--coordinate of the head groups in a specific leaflet for different compression rates and present these results in Fig.~\ref{fig-2}b \& c. For a relatively slow compression rate (8\e{-5} nm/ps), stalk growth follows a step-like process while for the higher compression rate (1\e{-4} nm/ps) it is more continuous. The relationship between mean curvature and area strain is the same at both compression rates in the absence of substrate defects. A relationship between the volume enclosed by a lipid protrusion and the compressive area strain of the bilayer has been recently proposed~\cite{Staykova2013}. Employing this relationship, the lipid stalks considered here will likely continue to elongate into tubes or tethers since the volume enclosed by the protrusions is minimal~\cite{Staykova2013} while the area strain is high. The negligible enclosed volume is an artifact that arises from the choice to use a substrate constructed from water molecules that are restrained from translation. 

\begin{wrapfigure}{r}{82.55mm}
\begin{center}
\vspace{-8mm}
\resizebox{0.46\columnwidth}{!} {\includegraphics{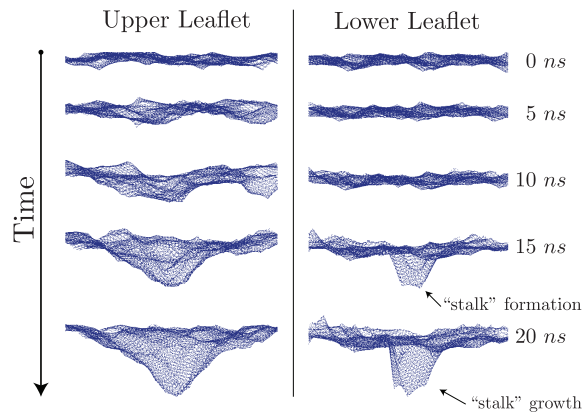}}
\end{center}\vspace{-10mm}
\captionsetup{font=footnotesize}
\caption{Views of the SLB along the simulation box diagonal. The upper (left) and lower (right) leaflet surfaces are shown every 5 nanoseconds. The images shown represent a section of each leaflet surface that spans the simulation box in order to observe the deformation near the center-defect in greater detail.} \vspace{-3mm}
\label{fig-3}
\end{wrapfigure}

Fig.~\ref{fig-3} shows the case where a 4 nm$^2$ hole in the supporting substrate is responsible for the primary mode of deformation. Snapshots of the upper and lower leaflets from the slowest compression rate simulation are displayed. When the lipid bilayer lies on a substrate that contains the defect, the primary response to compression is extrusion of the lower leaflet through the center defect. Simultaneously, the upper leaflet begins to curve toward the center of the defect. The extrusion process produces a more rapid increase in curvature with increasing area strain than is displayed for the defect-free system. The mode of deformation transitions from one that is center--defect controlled to being substrate protrusion controlled with increasing compression rate.  After this initial extrusion, buckling occurs at the corners of the simulation box in a manner similar to that for a defect-free substrate. However, the compression rate has a significant influence on the deformation mode that the SLB assumes. 

\begin{figure}[t]
\resizebox{1\columnwidth}{!} {\includegraphics{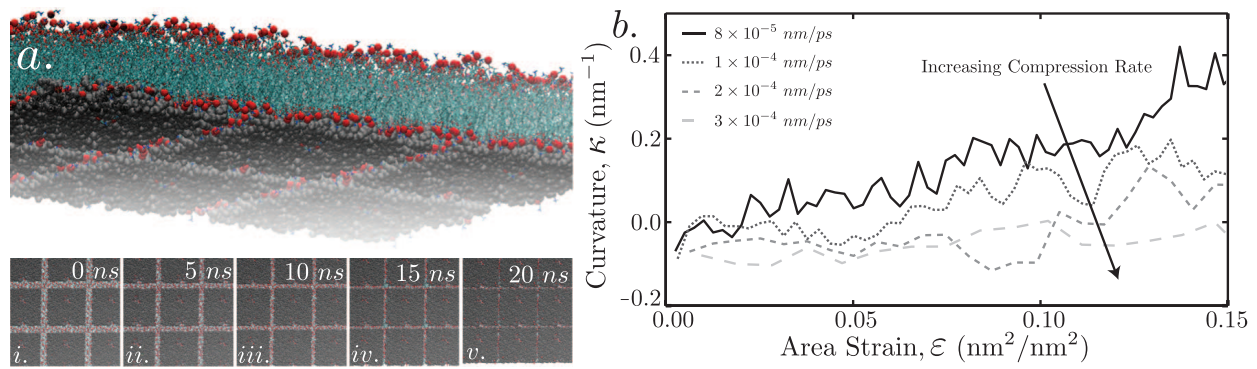}}
\captionsetup{font=footnotesize}
\vspace{-4mm}
\caption[]{($a.$) Image displaying the edge defect simulation case just prior to simulation.  The phosphorus atom in each lipid head group has been shown in red to better highlight the bilayer structure.  Hydrocarbon tails are displayed in blue.  Water molecules have been omitted for clarity.  (b) Bottom view of the simulation box and its 8 neighboring periodic images.  The images are snapshots displaying the system at $i.$ 0, $ii.$ 5, $iii.$ 10, $iv.$ 15, and $v.$ 20 ns.  The reduction in spacing between the substrate squares (grey) is quite prominent. ($b.$) Mean curvature (nm$^{-1}$) versus compressive area strain plot of the SLB lower leaflet. The data have been time averaged every 200 ps for the simulations containing a 4 nm$^2$ hole in the center of the supporting substrate.
\label{fig-4}}
\end{figure}

At the highest compression rate, the substrate protrusions control the deformation mode in a manner that is indistinguishable from the behavior of a bilayer compressed on a substrate without a center defect. Two possible explanations for this behavior are: (1.) the compression rate is faster than the rate the lipid membrane can adjust to the strain, or (2.) The substrate deformation at the corners of the simulation box is large enough to bias the system response toward those defects instead of the hole in the support. The mean curvature of the lower leaflet is plotted against membrane area strain for different compression rates in Fig.~\ref{fig-4}b.   Next, we consider the edge-defect scenario. The topology of the edge-defect substrate is analogous to supporting the bilayer with evenly spaced pillars. The pillars sit on a surface that undergoes equal bending about both horizontal axes. As bending increases, the separation distance between the tops of the supporting pillars decreases. Eventually these pillar tops join one another, recreating a continuous substrate. The membrane compression causes lipid molecules to extrude through portions of the open space surrounding the substrate. The bilayer deforms preferentially through the spaces at the corners of the substrate. During compressive biaxial loading, nanoscale substrate defects produce nontrivial changes in SLB deformation compared to a defect-free substrate. Such nanoscale defects could be present even in carefully fabricated substrates. This suggests that a defect-free substrate model may be invalid even for microscale SLBs.


\section{Conclusions}
In summary, we have examined the influence of specific nanoscale substrate defects on the behavior of SLBs under compression. The deformation of the lower leaflet of a bilayer induces lipid extrusion when a hole is included in the membrane support. At higher compression rates, the influence of a nanoscale hole on membrane deformation is greatly reduced. Whether larger defects require higher compression rates to mitigate their effects on membrane deformation remains unclear. 

While the experiments of Staykova {\em et al.}~\cite{Staykova2011} showed that lipid tethers form at the same location after repeated application and removal of biaxial compression to a SLB, the reason for this behavior remains unknown. Our results suggest that substrate defects serve as lipid tether nucleation sites, which provides a plausible explanation. The nontrivial changes in the dynamic behavior of SLBs due to nanoscale defects in the substrate show that assuming a defect--free support may not be advisable. Even small defects are capable of inducing changes in the deformation behavior of SLBs. Utilizing defects of nanoscale dimension opens up possibilities for creating novel 3D lipid structures by compressing bilayers on patterned membrane supports.

\section{Acknowledgements}

We thank the ESM Department Linux Computer Cluster and the Virginia Tech ARC SysX for providing us with computational resources.

\end{document}